# I See You: Teacher Analytics with GPT-4 Vision-Powered Observational Assessment

Unggi Lee[1], Yeil Jeong[2], Junbo Koh[2], Gyuri Byun[3]*, Yunseo Lee[4]*, Hyunwoong Lee[2], Seunmin EUN[2], Jewoong Moon[5] †, Cheolil Lim[2] † & Hyeoncheol Kim[1] †

Korea University, Department of Computer Science and Engineering, South Korea[1]
Seoul National University, Department of Education, South Korea[2]
Dangsu Elementary School, South Korea[3]
Pungnap Elementary School, South Korea[4]
The University of Alabama, Department of Educational Leadership, Policy, and Technology Studies, U.S.[5]

codingchild@korea.ac.kr / yell001@snu.ac.kr / gtkobo92@snu.ac.kr / gyuribyun97@gmail.com /
lyseo9772@gmail.com / leehw@snu.ac.kr / english0217@snu.ac.kr / jmoon19@ua.edu / chlim@snu.ac.kr /
harrykim@korea.ac.kr

† Corresponding Authors, *Equal Contribution

**ABSTRACT:** This preliminary study explores the integration of GPT-4 Vision (GPT-4V) technology into teacher analytics, focusing on its applicability in observational assessment to enhance reflective teaching practice. This research is grounded in developing a Video-based Automatic Assessment System (VidAAS) empowered by GPT-4V. Our approach aims to revolutionize teachers' assessment of students' practices by leveraging Generative Artificial Intelligence (GenAI) to offer detailed insights into classroom dynamics. Our research methodology encompasses a comprehensive literature review, prototype development of the VidAAS, and usability testing with in-service teachers. The study findings provide future research avenues for VidAAS design, implementation, and integration in teacher analytics, underscoring the potential of GPT-4V to provide real-time, scalable feedback and a deeper understanding of the classroom.

## 1. Introduction

The role of data-driven insights in promoting teaching practices has become increasingly essential (Jørnø & Gynther, 2018). Teacher analytics—a learning analytics field at the intersection of learning assessment and data analytics—has emerged as a critical contributor to this innovation (Camacho et al., 2020; Ndukwe & Daniel, 2020). It enables educators to harness various data types to refine their teaching approaches with more concrete and unobtrusive student data (Snodgrass Rangel et al., 2015). In contrast, existing methods of teacher assessment—primarily relying on classroom observations and student feedback—face limitations in capturing real-time, objective, and nuanced data of students' classroom interactions or learning practices.

The advent of emerging technologies—notably Generative Artificial Intelligence (GenAI)—opens a new avenue for vision and language processing, offering innovative solutions to longstanding challenges in educational assessments (Mao et al., 2023). For example, GPT-4 Vision (GPT-4V) can provide capabilities that outperform text-based analytics for the following reasons. By gathering, analyzing, and synthesizing multimodal data from various education settings, this approach is believed to enhance the understanding of classroom dynamics (Holstein et al., 2019; Martinez-Maldonado et al., 2022) and to refine the methods for observational assessments in educational settings.

In this study, we unveil the development and conceptualization of a Video-based Automatic Assessment System (VidAAS) seamlessly integrated with GPT-4V) (OpenAI, 2023b). This system mainly focuses on elevating the standards of observational assessment in educational contexts (Li et al., 2023). Our VidAAS targets leveraging the sophisticated image and video analysis capabilities of GPT-4V to deliver nuanced and data-driven feedback to educators. This approach—deeply rooted in teachers' reflective practice and cutting-edge technology—is aimed at enhancing the integration of teacher analytics.

Despite its promising perspectives on this technology, integrating VidAAS into teacher analytics is nascent. Various questions arise regarding the VidAAS's performance and strengths with challenges in real-world educational settings, which are crucial in its proper integration in schools. Our study is to identify actionable insights that can profoundly strengthen our understanding of both teaching methodologies and student learning processes. Accordingly, exploring the design of VidAAS and its potential synergy to educational practices is warranted to mark a significant stride towards technology-enhanced educational assessment. To guide our investigation, we hence formulated three research questions.

- RQ1. How can VidAAS web applications with GPT-4V be designed and developed for observational assessment?
- RQ2. What are the strengths, weaknesses, opportunities, and threats (SWOT) associated with implementing VidAAS utilizing GPT-4V in educational settings, and how do these factors influence its overall utility and adoption?
- RQ3. What are theoretical and practical frameworks supporting the effective integration of GPT-4V in a VidAAS, considering ethical guidelines, pedagogical effectiveness, and teacher professional development?

## 2. Literature Review

### 2.1. Teacher Analytics and Reflective Practice

Research indicates that technology-enhanced assessment and its design can empower teaching and learning (Pellegrino, 2006). Generally speaking, assessment aims to support complex class analysis and improve its efficacy by leveraging various technologies (Brown et al., 2008). Accordingly, there is an emerging need for exploring various technology-enhanced assessment approaches that effectively steer learning systems (Pellegrino, 2023). Of various assessment approaches, integrating automatic assessment systems (AAS) in various educational settings has transformed teacher analytics, fundamentally changing the ways educators reflect on their teaching and its practices (Worsley et al., 2021). In particular, these systems, especially in areas like affect-aware systems (Grawemeyer et al., 2017), educational data mining (Olugbade et al., 2018), and artificial intelligence-powered assessments (Ouyang et al., 2022), have increasingly provided the depth and breadth of educational data that can inform teachers' reflective practices and maximize student learning. Teacher analytics aims to enable detailed, objective, and data-driven decision-making. It allows educators to gain deeper insights into their teaching effectiveness and students' learning outcomes.

Reflective practice is deeply intertwined with teacher analytics. Reflective practice is when educators critically assess their teaching methods and student interactions (Alzoubi, 2022). Drawing from Donald Schön's theory on reflective practices (Griffiths & Tann, 1992; Schön, 1992), two aligned concepts guide this goal: (a) reflection-in-action and (b) reflection-on-action. Reflection-in-action is a process where an individual reflects on their action when it is occurring. In the context of teaching, it is related to a teacher's thinking about what they are doing while teaching. On the other hand, reflection-on-action aims to explore what one has done and consider how it could be improved in the future. In other words, it is closely related to thorough post-teaching reflections and long-term strategy planning (Loughran, 2002).

Both forms of reflection are vital in professional teaching and adaptive learning. Reflection-in-action allows educators to adapt to students' immediate needs and make real-time adjustments to their teaching approaches (Erdemir & Yeşilçınar, 2021). Conversely, reflection-on-action offers an opportunity for more in-depth and comprehensive analysis for long-term improvement (Nisson & Karlsson, 2019). Together, these reflective processes enable teachers to evolve their teaching strategies continually, enhance student learning, and respond effectively to the diverse needs of their classrooms (Slade et al., 2019). In this context, AAS can significantly enhance reflection-in-action when integrated into the teaching process. These systems provide real-time data and feedback about student learning trajectories (Kubsch et al., 2022; Moon et al., 2023), affective states (Dai & Ke, 2022), and performance (Slater & Baker, 2019). The synergy between AAS and reflective practices (in-action and on-action) creates a powerful tool for continuous improvement in teaching and student learning. Research has extensively involved AAS to evaluate performance, provide interactive feedback, and deliver detailed learner assessment results (Ihantola et al., 2010; Marchisio et al., 2018). For teacher analytics, incorporating AAS into this reflective cycle empowers educators with concrete evidence to support their reflections (Bannigan & Moores, 2009). This evidence-based approach to reflection enables teachers to make more informed decisions about their instructional strategies, classroom management, and curriculum design. Despite its potential, existing AAS might still be limited in fully grasping the nuances and contexts of teaching and student learning (Wang et al., 2022). Current AAS may not effectively capture the nuanced dynamics of learning or teaching-related actions. They struggle to capture, interpret, and analyze critical learning indicators automatically. Furthermore, existing AAS have difficulty understanding non-verbal cues, which are significant in teaching-learning. While teachers can intuitively interpret these signals, AAS still confronts challenges in recognizing the subtleties of non-verbal communication.

## 2.2. Observational Assessment: Significance and Challenges

Observational assessment—often guided by predefined checklists or rating systems and conducted by professional educators—is pivotal in evaluating learners' behaviors within educational contexts (Deno, 1985; Greenwood et al., 1994). Notable researchers and educators have extensively endorsed this method's ability to observe, listen to, and contemplate children's behaviors and utterances in real-time (Dunphy, 2010). It provides educators with valuable insights for data-driven decision-making (Peterson & Elam, 2020)

When conducting observational assessments, factors—such as scalability, accuracy, and a nuanced approach to complete assessment goals—are critical for objectivity and comprehensiveness (Walsh & Wolfgang, 2011; Marzano, 2010). Assessment contexts, including classroom environments and subject matters, demonstrate their critical roles in determining the appropriate scale and methods for observation. For example, leveraging technology can enable assessments to scale to larger classes or student groups while maintaining detailed observation capabilities. Observer bias can be another consideration because it may impact the accuracy of assessments. Minimizing this bias includes the application of standardized observation protocols empowered by AI-driven analysis tools to provide a more objective perspective. In terms of nuanced delivery, assessments should be aligned with specific learning goals and outcomes of the courses—ensuring that observations align with and accurately measure intended educational objectives. Ensuring reliability and validity is also inevitable, requiring well-defined criteria, consensus among raters, and monitoring of rating uniformity (Cone, 1982; Herman, 1992).

In observational assessments, exploring and validating new initiatives generally necessitates thoroughly examining their strengths, potential areas for improvement, and potential challenges. This comprehensive evaluation is vital for ensuring the reliability and validity of these methods (Cone, 1982). Teachers should select observational strategies that effectively evaluate specific learning-related aspects (Herman, 1992). Establishing well-defined criteria and specific rubrics—coupled with a unified understanding among evaluators about these criteria is essential to a consensus among raters about the interpretation of the criteria.

Exploring these methods brings attention to their practical application and effectiveness in natural classroom settings. This direction generally involves evaluating how user-friendly and adaptable these assessment tools are—ensuring they reach performance requirements. Furthermore, assessments need to explore areas for improvement and uncover opportunities associated with each approach. Technological advancements, such as video and voice recordings, have reshaped conventional techniques, allowing for more in-depth classroom monitoring and reflective analysis (Brandt & Perkins, 1973; Halle & Sindelar, 1982). Educators can use sophisticated evaluation methods combined with technology to assess the accuracy and effectiveness of tools in facilitating adaptive and diverse intervention design.

Recent interest among stakeholders in adopting various technologies is primarily centered on two goals: (1) gaining a deeper understanding of students' natural learning process and (2) evaluating the quality and effectiveness of teaching. Integrating technology in observational assessments can assist teachers in uncovering factors that influence students' behaviors and reflect on their classroom management strategies in the long term (Greenwood et al., 1994). This approach is aligned with current research—suggesting the advantages of technology-enhanced observational assessments that significantly improve teachers' insights into classroom dynamics. Pellegrino and Quellmalz (2010) thoroughly explored how technology may transform observational assessments—introducing more efficient and dynamic evaluation methods. Similarly, Shermis and DiVesta (2011) emphasized that instructors should possess a sense of ease and confidence when utilizing technology such as computerized essay grading or electronic portfolios. Vera and Castilleja (2016) also emphasized that digitized portfolios, such as digital anecdotal records and video files with document records, can structuralize and share the observation records.

## 2.3. Vision Language Model in Educational Research

The advancement of Vision Language Models (VLMs) has marked a significant milestone in deep learning—merging language and visual processing capabilities. VLMs, such as CLIP (Radford et al., 2021), Flamingo (Alayrac et al., 2022), BLIP-2 (Li et al., 2023), and LLaVA (Liu et al., 2023a, 2023b) have evidenced this wave in multimodal data analytics. CLIP utilizes unstructured text and a vast dataset for zero-shot learning, enabling accurate predictions across various domains. Flamingo improves upon VLMs with few-shot learning capabilities, integrating a frozen vision encoder and language model with cross-attention layers. BLIP-2 introduces the Query Former (Q-Former) to improve image-text matching and generation, keeping image encoders and Large Language Models (LLMs) frozen for peak efficiency. LLaVA (Liu et al., 2023a; 2023b) integrated a vision encoder with an LLM, offering a distinct simplicity in VLMs. Building upon this language model development, OpenAI recently introduced GPT-4V (OpenAI, 2023b; OpenAI, 2023c), making its application programming interface (API) accessible to the public. This innovation has taken a significant stride, particularly in the learning technology field, yet exploring educational research and practices remains relatively unexplored (Singh, 2023). In particular, investigating the capabilities of VLMs in this area was scarce. Despite a recent study by Lee and Zhai (2023) utilizing GPT-4V for automated scoring, the full potential of VLMs in educational research remains largely untapped. Accordingly, the proposed system in our study leverages the multimodal capabilities of GPT-4V to enhance the accuracy and autonomy of observational assessments in educational settings.

## 3. Method

The primary goal of this study is to explore and validate the integration of advanced generative AI (GenAI) technologies, particularly GPT-4V, in educational assessment. We aim to develop, explore, and test the effectiveness of a VidAAS that leverages the capabilities of GPT-4V for observational assessment.

### 3.1. Research Procedure

This research involves six design and development phases. The first phase was an extensive literature review that identified ideas of teacher analytics and reflective practice, observational assessment, and VLMs. In the second phase, the outcomes of this literature review were synthesized to yield the prototype design of the VidAAS web application. The third phase was conducting usability tests, where five teachers were interviewed post-use. The interview results were analyzed using a qualitative coding method in the fourth phase. Subsequently, the fifth phase employed a SWOT analysis, drawing upon the literature review and the coded interview results. The final phase culminated in the proposal of a comprehensive framework for VidAAS. This framework, informed by findings and developments from all previous phases, aims to guide future improvements in VidAAS, ensuring its effectiveness and relevance in academic support.

### 3.2. Participants

We interviewed subject matter experts (SMEs) in elementary education to investigate the usability, strengths, weaknesses, and perspectives on teacher analytics. The SMEs were selected based on their observational assessment proficiency and familiarity with AI technologies. Their profiles are detailed in Table 1 below.

*Table 1. The Information of the Interviewers*

| Name | Occupation | Major | Degree | Career (year) |
|---|---|---|---|---|
| Expert 1 | Elementary School | Elementary Education | B.Ed. | 8 |
| Expert 2 | | AI Education | M.A. | 8 |
| Expert 3 | | Educational Technology | M.A. | 9 |
| Expert 4 | | Elementary Education | B.Ed. | 9 |
| Expert 5 | Elementary School Teacher (P.E.) | Elementary Education | B.Ed. | 8 |

### 3.3. Data Collection and Data Analysis

We systematically gathered data through semi-structured interviews with five teachers. These interviews were designed to examine practical insights and implications of VidAAS. For data analysis, we employed both a thematic analysis and a SWOT analysis.

For thematic analysis, we conducted qualitative coding to extract meaningful insights systematically. First, we used data preprocessing to refine and familiarize data with data from a comprehensive transcript review. This phase was followed by the subsequent thematic analysis aligned with the structured coding approach by Corbin & Strauss (1990) that involves open, axial, and thematic coding. We then compiled a codebook delineating themes, codes, definitions, and examples. To further validate our analytical framework, we implemented in-vivo coding, categorizing feedback, and exploratory findings to reinforce the robustness of the present study.

The SWOT analysis involved a detailed examination of VidAAS's strengths, such as its potential to enhance observational assessment from a teacher analytics perspective. This analysis also explored the technology's weaknesses, including potential technological concerns over accuracy and reliability. Furthermore, it scrutinized the opportunities VidAAS presents in effectively transforming traditional pedagogical approaches. Lastly, the analysis addressed the potential threats to using VidAAS, involving ethical considerations, data privacy issues, and possible resistance to adopting innovative technologies in education. This multifaceted exploration aimed to convey a balanced view of the implications of integrating VidAAS into authentic learning environments.

## 4. Result

### 4.1. RQ1. How can VidAAS web applications with GPT-4V be designed and developed for observational assessment?

After reviewing current literature reviews and cutting-edge technology updates, we designed the prototype of VidAAS. This prototype features two primary components: (1) a web-based application and (2) a video-prompt pair sheet.

#### *4.1.1. Web-based Application*

#### *4.1.1.1. Application architecture*

We have developed the web-based application of VidAAS using advanced deep-learning models and Python libraries. The VidAAS employs video language models (VLMs) to understand and evaluate video content. The architecture unfolds in three stages: 1) *Video Processing*, 2) *Evaluation*, and 3) *User Interface Design*.
First, *Video Processing* is critical as it involves interpreting the visual and auditory information from the original video. This process unfolds two main parts; (1-1) *Converting Video to text*, and (1-2) *Audio to text*. In *Video to text* data, we harness the capabilities of the OpenAI API to access GPT-4V. Given that a video is essentially a sequence of images, GPT-4V is meticulously designed to read each frame. We have developed a specialized script that breaks the video into its frames, allowing GPT-4V to analyze them individually. This methodological approach enables the generation of a text description that captures the essence and content of the video comprehensively. In parallel, the *Audio to text* plays a critical role in ensuring understanding of the video content. Here, we leverage OpenAI's Whisper AI, a tool renowned for reasonably accurately converting spoken words into text. Whisper API provides a highly-quality transcription of the spoken content by listening to the video's audio track. This complements the visual analysis generated by GPT-4V and ensures that the nuances and subtleties delivered into audio are not lost, resulting in a rich and multi-dimensional understanding of the video content. Together, these processes lay a solid basis for further analysis and evaluation, corroborating that everything is noticed in the quest to provide a thorough assessment of the video material.

Second, *Evaluation* is vital for assessing the video's content. This stage is designed to utilize rubric-based prompts and the capabilities of GPT-4 to perform a thorough evaluation. It unfolds through a three-part process involving (2-1) *Prompt chaining process*, (2-2) *Integration with LangChain*, and (2-3) *Evaluate with GPT-4*. First, the *Prompt chaining process* is a technique where the system's operations are structured through a 'prompting chain' method. This technique guides the LLMs through a step-by-step process. Building upon each prompt allows the AI to thoroughly process the video and audio descriptions and generate coherent results that are closely aligned with the assessment rubrics. Second, *Integration with LangChain* is essential to facilitate and streamline the prompt chaining process. *LangChain* is a pivotal tool that facilitates the prompt chaining process. It refines the interaction with language models by automating the creation of prompt sequences and managing the flow of these operations. This stage not only simplifies the AI's understanding and execution but also ensures higher accuracy and efficiency in the evaluation process. Third, *Evaluate with GPT-4* is the stage that aims to leverage the power of GPT-4 for evaluation. Building on previous research (Lee et al., 2023a), we utilize GPT-4's advanced capabilities for our evaluation purposes. GPT-4 receives inputs comprising the fragmented evaluations of the video. It then employs the assessment rubrics as integrated prompts to synthesize these fragments. GPT-4 can effectively summarize the evaluations and draw conclusive insights through this step. This approach ensures that the final evaluation is comprehensive and nuanced, reflecting a deep understanding of the video's content.

Third, *User Interface Design* is a step where the web-based application becomes accessible to users. This stage involves two parts: (3-1) *User interface with Gradio* and (3-2) *Web application hosting via Hugging Face Space*. First, we start with *Gradio*, a Python library renowned for transforming our complex machine learning (ML) code into user-friendly, interactive web applications. This step helps users to engage with our application effortlessly, enjoying a seamless and intuitive experience. Second, *Web Application Hosting via Hugging Face Spaces* is to deploy our Gradio web interface to users. Hugging Face Spaces provides a robust platform offering the necessary infrastructure to make our web application publicly accessible. Figure 1 is the whole architecture of the VidAAS web application. This step ensures that users can easily access our application, regardless of their location or device, with minimal setup required. Table 2 involves the details of VidAAS deployment features.

*Figure 1. Whole Architecture of VidAAS Web Application*

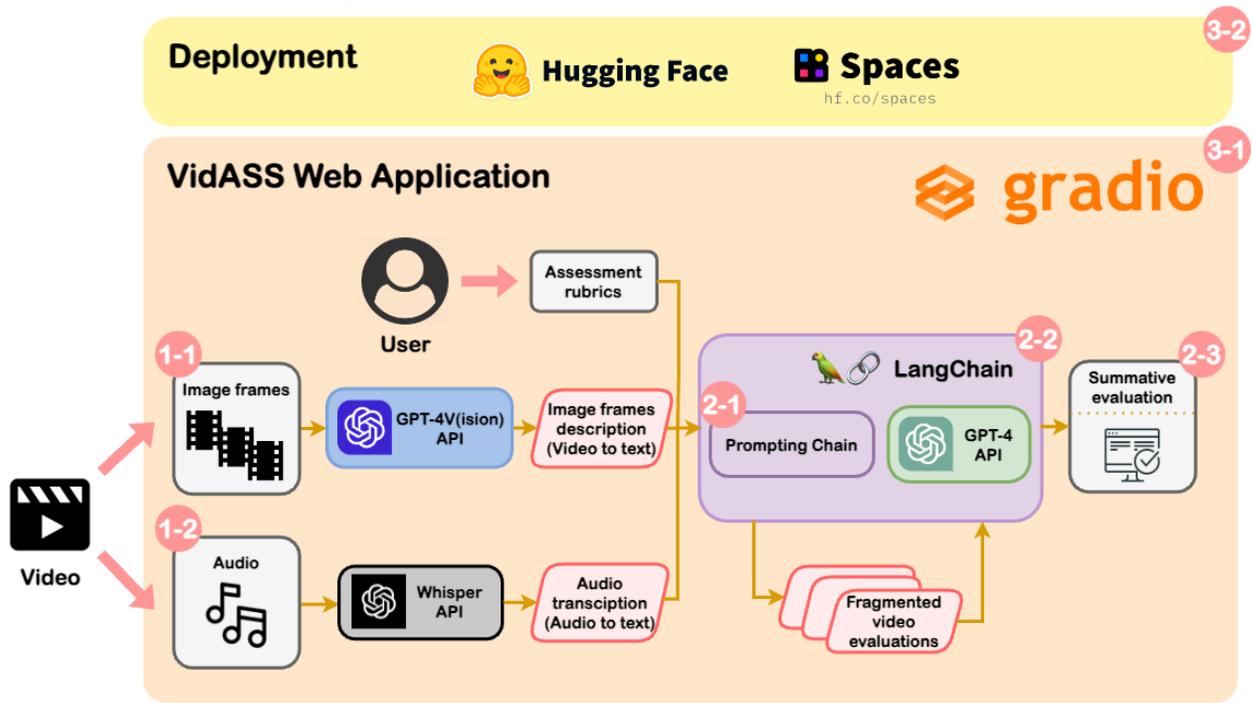

*Table 1. The Information of the Interviewers*

| Technology | Brief explanation |
|---|---|
| GPT-4V API | The GPT-4V API, offered by OpenAI, enables developers to create their products by utilizing this API. We used "gpt-4-vision-preview", a beta version of GPT-4V API, because OpenAI does not serve alpha versions of GPT-4V API. The API functionality fully integrates with our VidAAS web application. |
| Whisper API | Whisper is an automatic speech recognition (ASR) AI model by OpenAI (OpenAI, 2022). We used Whisper by calling API requests in the same way as GPT-4V. |
| LangChain | LangChain is a Python library designed to enhance the capabilities of LLMs by providing tools for efficient prompt engineering and chaining. |
| Gradio | Gradio is an open-source Python library for creating interactive ML web applications. It allows developers to build and share interfaces for their ML models quickly. |
| Hugging Face Space | VidAAS is deployed in Hugging Face Space. Hugging Face Space is a platform that offers a simple way to host ML applications. It also provides built-in support for Gradio SDKs, letting users build applications in minutes. |

*4.1.1.2. Steps to use*

The VidAAS web application begins with the first step: 'Make batched snapshots & audio script.' As shown in Figure 2, users can upload their video and receive displayed frames and audio transcriptions via OpenAI Whisper. To balance the performance efficiency and cost of the app, users can customize the number of image frames used for evaluation, optimizing the trade-off between performance and resource use. This step sets the foundation for a tailored and efficient user experience.

*Figure 2. First Step of VidAAS Web Application*

**GPT-4 Vision for Evaluation**

**1st STEP. Make Batched Snapshots & Audio Script**

Enter your OpenAI API Key
Your API Key must be allowed to use GPT-4 Vision

sk-*********...

Upload your video (under 1 minute video is the best..!)

파일을 끌어 놓으세요
- 또는 -
클릭해서 업로드하기

Number of images in one batch          5
Choose between 2 and 5

Percentage(%) of batched image frames to total frames     3
Choose between 1(%) and 5(%)

Process

Batched Snapshots of Video

Audio Transcript
(You can EDIT by yourself)

The second step of the application (Figure 3) is 'Set evaluation rubric.' In this step, users choose the types of information for analysis— 'Video + Audio' or 'Video only.' Following this selection, they define the video evaluation rubric. Once set, users click the 'Evaluate' button to initiate the analysis of video content based on the chosen rubric, moving the process forward to the evaluation stage.

The third step of the application (Figure 4) is 'Summarize and get results.' Given the challenge of understanding extensive analysis from GPT-4, this step provides users with a concise summary of the information and scores based on the evaluation rubric. We employ GPT-4 and prompt chaining techniques, ensuring users receive a clear and manageable overview of the results.

### 4.1.2. Video-Prompt Pair Sheet

To conduct usability tests and demonstrate the effectiveness of VidAAS, we have created video-prompt pair sheets. These pairs include videos selected for observational assessment and corresponding rubric prompts tailored to each video. Researchers carefully selected videos needed for observational assessment and assessment rubric prompts matching the videos. Details of the video-prompt pairs are provided in Appendix 1, allowing users to understand the assessment criteria and the types of content evaluated.

*Figure 3. Second Step of VidAAS Web Application*

*Figure 4. Third Step of VidAAS Web Application*

**4.2. RQ2. What are the strengths, weaknesses, opportunities, and threats (SWOT) associated with implementing VidAAS utilizing GPT-4V in educational settings, and how do these factors influence its overall utility and adoption?**

*4.2.1. Usability Test*

To test VidAAS's usability, we conducted semi-structured interviews with the five teachers (see Participant section). The usability test focuses on 1) the advantages and disadvantages of VidAAS's current usability, 2) Potential benefits and areas to be improved, 3) teacher analytics considerations, 4) criteria for effective and ethical use, and 5) varied views on human-AI collaboration.

*4.2.1.1. Advantages and disadvantages of VidAAS's current usability*

The primary benefits of the VidAAS tool involve its high accuracy in evaluating skills in the behavioral (psychomotor) domain. In addition, it offers comprehensive explanations for each assessment, aligning with specific rubric criteria. Nevertheless, the VidAAS tool faces several design challenges. It is constrained by video length and processing speed, hindering real-time assessment. Scaling to handle large volumes of assessments requires efficiency improvements. In addition, enhancing the accuracy of its video recognition is crucial for better performance.

*4.2.1.2. Potential benefits and areas to be improved*

The interviewees noted that teachers have the potential to better observe and evaluate students' performance in authentic contexts. VidAAS has the potential to serve as a meticulous observation assistant and an unbiased decision-making aid in authentic learning settings, enhancing the accuracy and objectivity of assessment. This approach could decrease evaluation anxiety, allowing for ongoing assessment over time rather than a single instance and giving students multiple opportunities to demonstrate their capabilities. In addition, VidAAS proves advantageous to teachers' more profound understanding of students' inherent learning processes.

Despite the system's promise, certain areas require enhancement to realize its potential benefits fully. To support process-oriented assessments, developing capabilities for evaluating learning progress over extended periods and analyzing improvements is vital. In Bloom's taxonomy, improvements are particularly needed in cognitive and affective domains. Enhancing the system's ability to comprehensively understand the context and identify external indicators for observations inferring individuals' cognitive activities are critical for further development.

*4.2.1.3. Teacher analytics considerations*

The interviewees also identified two critical applications for 'reflection-in-action' and 'reflection-on-action' within teacher analytics using VidAAS. First, it can be an in-class learning analytics tool for real-time, process-oriented assessment, potentially enhancing teachers' profession through these data-driven practices. Moreover, novice teachers could benefit by reviewing their lessons to identify overlooked aspects. Also, they can retrospectively analyze their students' performance and teaching techniques together. This could facilitate data-driven decision-making of novice teachers.

However, there are concerns regarding the system's limitations in promoting teachers' professional development. One interviewee noted that while VidAAS could be an effective feedback tool for student improvement, its capacity to understand qualitative aspects, such as human interactions in classroom settings, appears limited. Hence, he thought the system might be limited in enhancing teachers' related skills.

*4.2.1.4. Criteria for effective and ethical use*

The interviewees noted that to effectively and ethically utilize VidAAS, specific requirements will be considered. First, teachers need to understand the significance of interpreting data within contexts. Second, users should align the tool's use and functions with students' developmental levels and set appropriate learning goals. Third, detailed usage processes and protocols for various purposes must be subsequently presented. Last, the suitable device for different usage intentions should be chosen carefully.

Interviews with the teachers also revealed vital design principles: User Experience, Design, and Functions. First, regarding user experiences, there are two principles to be addressed: 1-1) Providing additional support in establishing rubrics, and 1-2) Enhancing accessibility across different platforms such as digital textbooks and smart glasses. Second, regarding design perspective, the following aspects are shown: 2-1) Creating distinct interfaces tailored to the specific needs of teachers and students, 2-2) Enabling the examination of long-term performance improvements through archived video footage and automated assessment records. Third, three principles will guide functions: 3-1) Reducing video processing latency for real-time analysis, 3-2) Enabling concurrent mass evaluation for multiple students, and 3-3) Support extensive video context for evaluating high-order thinking.

Finally, adhering to ethical guidelines is paramount. Teachers noted that users need to (1) anonymize personal information before using the tool (Converting data to unidentifiable one), (2) robust data security measures, and (3) carefully determine consent for the use of personal information. By actively addressing these considerations, stakeholders can maximize VidAAS's potential while maintaining ethical integrity and focusing on user-centric design.

*4.2.1.5. Varied views on human-AI collaboration*

Interviewees displayed diverse attitudes toward AI, perspectives on human-AI collaboration, and philosophies on educational assessment. First, attitudes toward AI varied widely. Some expressed positive attitudes, satisfied with the current capabilities and optimistic about future advancements. Others held negative or limited views, citing AI's inability to emulate human qualities or behaviors. Second, opinions on AI-human collaboration also vary. The perspective shift from viewing AI merely as a tool to seeing it as a potential collaborative partner was notable. Some suggested that AI could serve as distributed cognition, offering guidance in areas beyond human capacity while underscoring that humans should retain final decision-making authority. Third, perceptions of AI-assisted assessments were mixed. While some appreciated AI's understanding of natural language rubrics and potential efficiencies, others distrusted AI, holding a skeptical view of its reliability and judgment. Table 3 below is a summarized version of these findings above. Details with related quotes and analyses are available in Appendix 2.

*Table 3. Findings from Qualitative Coding*

| Themes | Codes | Implications |
|---|---|---|
| 1. Usability of current VidAAS tool | 1.1. Advantages | 1.1.1. High rate of correct evaluation in the behavioral (psychomotor) domain |
| | | 1.1.2. Comprehensive explanation for the evaluation |
| | 1.2. Disadvantages | 1.2.1. Limited video length and video processing latency |
| | | 1.2.2. Unfeasible massive amount of assessment |
| | | 1.2.3. Low rate of recognition precision |
| 2. Potential benefits and improvement points of future VidAAS tool | 2.1. Benefits | 2.1.1. Potential of observing and evaluating in authentic contexts |
| | | 2.1.2. Potential to mitigate assessment anxiety |
| | | 2.1.3. Potential to enhance the range of evaluation methodologies |
| | | 2.1.4. Potential to enhance evaluation accuracy and objectivity(fairness) |
| | | 2.1.5. Potential to comprehend students' innate learning process with real-time analysis |
|  |  | 2.1.7. Potential to provide feedback, scaffoldings and encouragement for students as a self-evaluation instrument |
|  | 2.2. Improvements | 2.2.1. Considering comprehensive evaluation of past cumulative observations |
|  |  | 2.2.2. The ability to consider context comprehensively to for the evaluation of cognitive and affective domain |
| 3. Considerations of usage in relation to teacher analytics | 3.1. Reflection in | 3.1.1. A tool for teachers to do learning analytics in class |
|  |  | 3.1.2. A tool to facilitate real-time, process-oriented assessment |
|  | 3.2. Reflection on | 3.2.1. Watching a recording and reflecting on the lesson |
|  |  | 3.2.2. Looking back at student's or teacher's performance |
|  |  | 3.2.3. Data-driven decision-making |
|  | 3.3. Doubts (Limitations) | 3.3.1. Undetailed understanding of qualitative factors |
| 4. Requirements for effective and ethical utilization | 4.1. Usage implications | 4.1.1. Informed of the need for contextualized interpretation of human teachers |
|  |  | 4.1.3. Contextualized usage process (protocol) in detail |
|  | 4.3. Ethical guidelines | 4.3.1. Replacement of identifiable personal information |
|  |  | 4.3.2. Thorough functional readiness for data security |
|  |  | 4.3.3. Consent to use of personal information |
| 5. Diverse perspectives regarding AI | 5.1. Attitudes toward AI | 5.1.1. Satisfaction with the current level and a positive perception for future advances |
|  |  | 5.1.2. Perspective of AI that it cannot be seen as a person |
|  | 5.2. AI-human collaboration | 5.2.1. Perspectives of AI (whether as a meer tool or a collaborative entity) |
|  |  | 5.2.2. AI as a distributed cognition |
|  |  | 5.3.2. Negative attitudes with a distrust of AI |

*4.2.2. SWOT Analysis*

SWOT is the analytical framework that distinguishes the strategies in the categories of strengths, weaknesses, opportunities, and threats (Henzaghta et al., 2021); the framework has been used for education research, specifically new technology's adoption and usability testing (Zhu & Justice Mugenyi, 2015). We employed the SWOT analysis methods to thoroughly investigate VidAAS's future design, delivery, and research directions. We present VidAAS's fundamental functions and usability, integrating theoretical insights from teacher analytics, AI, vision models, and general educational applications. This analysis revealed the system's internal strengths and weaknesses, focusing on its educational utility and usability and external opportunities and threats based on its broader educational impacts. Table 4 is the summarization of the SWOT analysis.

VidAAS's key strengths include its ability to provide objective, rubric-based video assessments, offering detailed evidence for logical and understandable evaluations. It could excel in quantitatively assessing various indicators, particularly affective domains. In addition, it offers in-depth analyses of students' performance processes, indicating their change in cognitive or psychomotor domains.

These strengths lead to several opportunities for VidAAS. First, it assists teachers in enhancing their skills through 'reflection-in-action' by analyzing and illustrating students' learning trajectories, suitable for scalable, real-time, and process-oriented assessments. Second, it supports 'reflection-on-action' for teachers, allowing them to refine their instruction based detailed assessment for their teaching, aiding particularly novice educators in data-driven decision-making (e.g., parent-teacher consultation). Third, students benefit by using VidAAS for feedback on self-directed learning, enhancing their abilities and competencies through different forms of assessments (i.e., peer- or self-assessment). Finally, VidAAS diversifies assessment methods, alleviating test anxiety and fostering a more natural demonstration of students' abilities.

However, VidAAS also presents some weaknesses. First, technical limitations impeded by current versions hinder its efficiency; video processing is time-consuming, and the accuracy of analysis is perceived variably by individuals, making scalable assessments challenging. Second, VidAAS lacks the necessary context sensitivity, unable to adjust its use to specific educational settings or user needs. Which often necessitates teachers to reinterpret data. In addition, it may struggle to assess certain affective domains and nuanced aspects of student-teacher interactions. Third, the black-box nature of current AI may obscure clear understanding of how assessments are derived, augmenting a skepticism on its ability to accurately reflect learning objectives. Last, ethical concerns arise from the potential risk regarding facial data storage without proper anonymization.

The identified weaknesses of current VidAAS pose specific threats for both students and teachers. First, employing VidAAS without adequate guidelines could confuse teachers, particularly those less proficient with technologies (Lim, 2023; Nazaretsky et al., 2022). This confusion is exacerbated by the recent significant shift of AI adoption that has been brought to teaching practices. Second, VidAAS may invade teachers' evaluative authority and controls. With the system providing comprehensive assessments, the reliance and trust in human judgment, particularly educational assessment, might diminish despite the growing importance of teachers' autonomy in student assessment (Varatharaj, 2018). Last, using VidAAS without proper ethical safeguard could compromise student privacy and generate ambiguity in data protection of assessment outcomes. These potential threats echo the need for careful and legitimate consideration in future VidAAS's implementation.

*Table 4. Result of SWOT Analysis*

| SWOT | Contents | Related implications from coding |
|---|---|---|
| Strengths | 1. Provides objective assessments | 2.1.4. 1.1.1. |
| | 2. Provides detailed assessments of students' performance processes | 1.1.2. 2.1.1. |
| Weaknesses | 1. Has limitations due to current state of technical capabilities | 1.2.1. 1.2.2. 1.2.3. |
| | 2. Not able to reflect the context of education in detail | 2.2.1. 2.2.2. 3.3.1. |
| | 3. Determines the assessment results only by the AI model | 4.1.1 5.1.2. |

|  |  |  |  |
|---|---|---|---|
|  | 4. | Concerns on the ethical issue | 4.3.1.<br>4.3.2.<br>4.3.3. |
| Opportunities | 1. | Improves teachers' teaching skills by analyzing students in class | 3.1.1.<br>3.1.2. |
|  | 2. | Reflects teachers' instruction | 3.2.1.<br>3.2.2.<br>3.2.3. |
|  | 3. | Enhances students' abilities by getting feedback | 2.1.5.<br>2.1.7. |
|  | 4. | Diversifies the assessment methods | 2.1.3. |
|  | 5. | More accurate assessment | 2.1.2. |
| Threats | 1. | Causes confusion to teachers | 4.1.3. |
|  | 2. | Infringes the teachers' evaluative authority | 5.3.2. |
|  | 3. | Leads to ethical issues | 5.2.2. |

### 4.3. RQ3. What are theoretical and practical frameworks supporting the effective integration of GPT-4V in a VidAAS, considering ethical guidelines, pedagogical effectiveness, and teacher professional development?

Based on our SWOT analysis findings, we proposed the VidAAS framework (Figure 5), envisioning a systematic approach to observational assessment with VLMs. The VidAAS framework is structured around three main elements: (1) *Goal*, (2) *Components*, and (3) *Availability*.

First, the *Goal* field portrays the framework's fundamental purpose. The VidAAS framework's goal is twofold. One aspect is supporting teachers' assessment performance; VidAAS aims to support teachers' observational assessments, complementing rather than replacing their role in the process. It implies that while VidAAS provides assistance, teachers maintain overall controls for the assessment process and fully consider the idea of 'human-in-the-loop'. The second aspect focuses on promoting teachers' reflective practices for professional development. VidAAS serves as a tool for reflection, helping teachers, especially novice, identify areas of improvement. Interview results corroborate that teachers view VidAAS as a beneficial resource for bolstering their reflective practices and professional growth.

Second, the *Components* field of the VidAAS framework addresses three aspects. *Intellectual Facilitation and Assistance* indicates that VidAAS acts as a cognitive aid that delivers timely support to both teachers and students via simultaneous and complementary assessments. Through automatic observational assessment, VidAAS informs scores and feedback to teachers immediately. Thus, VidAAS can play a role in being an intellectual facilitator, augmenting the teaching and learning processes. In addition, VidAAS prompts teachers to reflect on their evaluation and selectively apply feedback, thereby improving their evaluation capabilities. Concurrently, students are encouraged to engage in self- and peer-assessment. Interview records reveal that students could utilize VidAAS as a formative assessment tool that monitors learning progress, effectively sharing cognitive resources, which was previously shouldered by teachers alone. This approach to *Contextualized Self-Improvement* suggests that the VidAAS system supports teachers' professional growth by fostering reflective practices. Specifically, VidAAS facilitates reflective practices by allowing teachers to track and compare their assessment records over time, yielding insights into further development of their teaching strategies and student performance. This longitudinal perspective offers teachers deeper insights of how their teaching methods and students' performances evolve. This feature enables teachers to understand the context of each assessment better, seeing not just a snapshot but a progression. Such a feature in the VidAAS is particularly advantageous to pre-service teachers because it offers them a structured guide that reflects on their teaching approaches. This feature enables users to track development and pinpoint areas requiring improvement, offering a clear and chronological perspective of their progress. *Multi-Dimensional Support* refers to enhancing the assessment process by minimizing bias and ensuring objective commentary. This is achieved by integrating features that promote multiple perspectives and support both reflection-in-action and reflection-on-action. Such an approach encourages multidimensional thinking and leverages the expertise of secondary raters. These raters can offer viewpoints, helping to fine-tune and validate assessments, thus enhancing the overall accuracy and depth of the assessment.

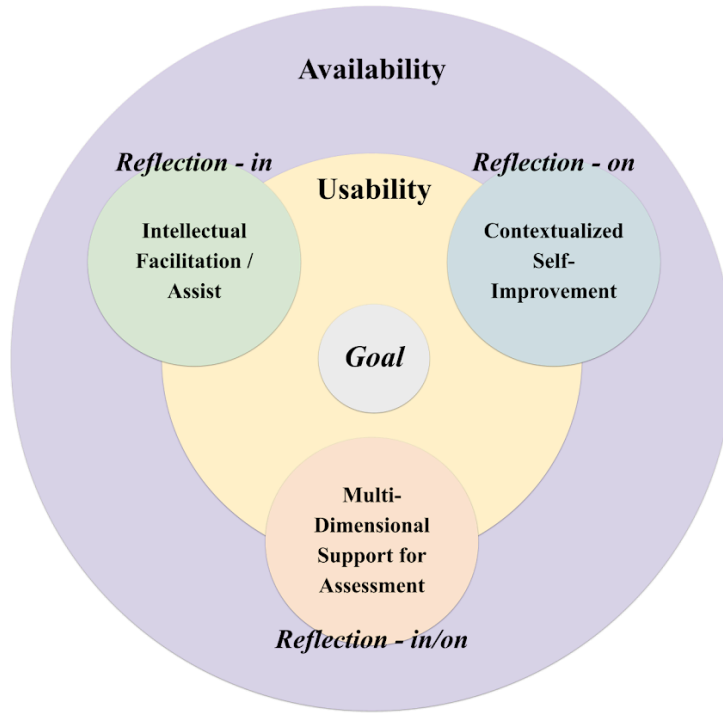

*Figure 5. VidAAS Framework*

| Goal | · To support teachers' assessment performance. <br> · To support teacher reflection to improve professional development. | | |
|---|---|---|---|
| Components | **Intellectual Facilitation / Assist** <br> · Cognitive supports with simultaneous assessments <br> · Providing scores and feedback to teachers <br> · Supports for self-assessment and peer assessment | **Contextualized Self- Improvement** <br> · Supporting teacher reflection <br> · Comparing assessments with contextualization <br> · Comparison of assessments in chronological order in a relative assessment approach <br> · Applicability to preservice teachers | **Multi-Dimensional Support** <br> · Minimizing bias and providing objective commentary <br> · Promoting multiple perspectives, multifaceted reflection, and multidimensional thinking <br> · Utilizing secondary raters to adjust and refine assessments |
| Accessibility | **Usability** <br> · When clear evaluation criteria such as rubrics exist <br> · When conducting mass observation evaluations in various areas <br> · When conducting an evidence-based evaluation of the performance process | **Availability** <br> · A permissive atmosphere in the community <br> · Consent to use personal video and audio | |

Third, the *Accessibility* field of the VidAAS framework centers on two main components: *Usability* and *Availability*. *Usability* is essential, especially when clear evaluation criteria, such as rubrics, are used for scalable, observational, and evidence-based performance assessments. This component requires a system to make intuitive and straightforward assessment that results in enhanced effectiveness and efficiency. On the other hand, *Availability* addresses fostering an environment of permission within a community. It necessitates obtaining consent for the use of personal video and audio materials, a vital step for upholding ethical standards and individual privacy. Obtaining such consent ensures that VidAAS operates effectively without violating personal rights or stimulating discomfort and distrust. Figure 5 presents the integrated structure of the VidAAS framework, encompassing these considerations.

## 5. Discussion

In this research, we explored the potential of VidAAS to support observational assessment in a lens of teacher analytics. After the thorough literature review, usability test, and the SWOT analysis, we yielded several implications and suggestions for the future design of VidAAS.

### 5.1. Implications of Study

The study findings give major implications for VidAAS's role in educational research and practices. First, VidAAS has shown promise as a reflective tool for educators. It supports real-time adaptation to student needs (reflection-in-action) (Nian, 2020; Schön, 1992) and retrospective lesson analysis (reflection-on-action) (Schön, 1992), offering a more efficient feedback mechanism compared to conventional methods (Pellegrino & Gerber, 2012; Tripp & Rich, 2012), which often involved inefficient and delayed approaches, such as manually reviewing video footage or requiring expert guidance, by offering a more streamlined and efficient way to convey feedback. The proposed system allows teachers, especially those new to the profession, to carefully review recorded lessons to identify areas of improvement and better understand effective teaching techniques. However, it is important to note that when grasping the full spectrum of classroom interactions, it is necessary to consider a balanced approach that combines VidAAS with human insights to fully grasp educational complexities.

Second, while VidAAS significantly aids educators, it is not a replacement for human judgment. Teachers, both in this and prior research (Lee et al., 2023; Han & Lu, 2021), have noted that their assessments go beyond strict rubric adherence to include subjective judgments, reflecting an understanding of individuals' progress and the learning contexts. This positions VidAAS as a supportive tool that augments, not replaces, teachers' observational assessments. In this collaborative context, the AI provides data-driven insights, which teachers then interpret and contextualize (Holstein et al., 2018; Kim & Cho, 2022; Kuniyoshi & Kurahashi, 2020). Integrating AI in educational practices should thus be a complementary process, combining the strengths of AI and human insights to improve educational experiences and outcomes (Ouyang & Jiao, 2021). This perspective underscores the ongoing need to evolve AI tools in education, aiming not to replace but enhance human capabilities, thereby illustrating the importance of integrating a 'human-in-the-loop' approach in the system.

Third, VidAAS offers pre-service teachers an opportunity for professional development. Many beginner teachers face a disconnect between classroom realities and unified theories, struggling to apply their knowledge and skills (Allen, 2009; Allen & Wright, 2014; Phillips & Condy, 2023) in real-world settings (Resch et al., 2022; Resch & Schrittesser, 2023). The proposed VidAAS can bridge this gap. By allowing them to review and analyze recorded lessons, pre-service teachers can gain insights, strategies, and peer feedback. This process not only refines their teaching profession and enhances overall knowledge and teaching techniques.

Last, incorporating VidAAS into teacher analytics (Worsley et al., 2021) presents a new avenue for learning analytics and educational data mining fields (Prieto et al., 2016). The field of learning analytics has widely focused on teacher analytics by exploring teaching integrations, student-teacher interactions, and data-driven decision making by technology-enhanced learning (Camacho et al., 2020). It has provided insights in how educators can adapt their teaching to individual student needs. The present study contributes to this field by introducing a GPT-4V powered observational assessment. It supports real-time and retrospective analysis, offering a more efficient and nuanced feedback mechanism. The study findings deliver how recent GenAI technologies boost the teacher analytic system and how it could transform conventional observational assessments. Future studies may open continuous discussion regarding ways to leverage AI approaches to enhance teacher analytics.

### 5.2. Suggestions for Future Research

There are areas for future research suggestions. First, enhancing VidAAS's speed and scalability is essential. Despite GPT-4V's promising performance, its current limitations in real-time assessments and managing multiple learners simultaneously are evident. Exploring and testing more efficient VLMs such as LLaVA (Liu et al., 2023a; 2023b) and Video-LLaVA (Lin et al., 2023), which are smaller than GPT-4V, could boost real-time processing capabilities. In addition, leveraging Semantic Analysis Models (SAM) proposed by Kirillov et al. (2023), along with visual prompt technologies (Jia et al., 2022; Wang et al., 2023), could offer effective ways to assess numerous students in a scalable manner.

Second, to enhance VidAAS's effectiveness, we have derived key design principles from a SWOT analysis, concentrating on user experience, design, and functions. For user experience, we aim to streamline the rubric setup with clear steps and examples and ensure compatibility with various educational technologies. From a design perspective, we plan to create distinct interfaces tailored for teachers and students, addressing their unique needs. This includes interfaces for teachers to manage multiple learners effectively and providing learners with both knowledge-base analytical feedback and motivational feedback. In addition, we aim to facilitate access to long-term performance data through archived footage and evaluation. Future studies may consider iterative development of real-time analysis with minimal latency, enabling simultaneous evaluations of multiple students, improving the assessment of cognitive and affective domains in multifaceted and complex contexts. These principles, detailed further in Appendix 3, guide the development of a more efficient and user-friendly VidAAS in future design cycles.

Third, this study highlighted the limitations of VidAAS in capturing qualitative nuances of classroom dynamics. Future research needs to seek ways to infuse qualitative feedback mechanisms into the system. Potential avenues could include developing algorithms capable of interpreting non-verbal cues and emotional states or mechanisms when integrating direct feedback from teachers and students (Abdulghafor et al., 2022; Zhang et al., 2020). Also, it is crucial to adapt VidAAS for various educational settings and individual learners' needs. Following studies will concentrate on delivering customization features that allow teachers to modify the system to fit their teaching approaches and student demographics. This could entail adaptive learning algorithms that adjust assessment criteria based on individual student progress and needs (Kabudi et al., 2021).

## 6. Conclusive Remark

In our research, with the lens of reflective practices, we developed the VidAAS web application, integrating GPT-4 Vision and cutting-edge technologies. Through usability testing with teachers, qualitative analysis, and SWOT analysis, we envisioned the current and future design and delivery of VidAAS. These insights, along with literature review findings, informed the development of the VidAAS framework to advance research in automatic observational assessment empowered by Generative AI and Vision Language Models (VLMs). However, the VidAAS web application faces scalability challenges and future design issues, particularly with the growing need toward mobile usage. In addition, the emerging and nascent nature of VLM technology indicates that the current system needs more enhancement through future research. The VidAAS framework requires comprehensive real-world testing to examine its effectiveness. Moving forward, our future goal is to explore and optimize the role of Generative AI in automatic observational assessment, particularly within the realm of teacher analytics.

# References

Abdulghafor, R., Abdelmohsen, A., Turaev, S., Ali, M. A. H., & Wani, S. (2022). An analysis of body language of patients using artificial intelligence. *Healthcare 2022, 10*, 2504. https://doi.org/10.3390/healthcare10122504

Ahmed, K., Miskovic, D., Darzi, A., Athanasiou, T., & Hanna, G. B. (2011). Observational tools for assessment of procedural skills: a systematic review. *The American Journal of Surgery, 202*(4), 469-480. https://doi.org/10.1016/j.amjsurg.2010.10.020

Allen, J. (2009). *The" theory-practice gap": Turning theory into practice in a pre-service teacher education program* (Doctoral dissertation, CQUniversity).

Allen, J. M., & Wright, S. E. (2014). Integrating theory and practice in the pre-service teacher education practicum. *Teachers and teaching, 20*(2), 136-151. https://doi.org/10.1080/13540602.2013.848568

Alayrac, J. B., Donahue, P. L, A. Miech, I. Barr, Y. Hasson, K. Lenc, A. Mensch, K. Millican, M. Reynolds, R. Ring, E. Rutherford, S. Cabi, T. Han, Z. Gong, S. Samangooei, M. Monteiro, J. Menick, S. Borgeaud, A. Brock, A. Nematzadeh, S. Sharifzadeh, M. Binkowski, R. Barreira, O. Vinyals, A. Zisserman, & K. Simonyan. (2022). Flamingo: a visual language model for few-shot learning. *Advances in Neural Information Processing Systems, 35*, 23716-23736. https://doi.org/10.48550/arXiv.2204.14198

AlZoubi, D. (2022, January). From Data to Actions: Unfolding Instructors' Sense-making and Reflective Practice with Classroom Analytics. *In Proceedings of 12th International Conference on Learning Analytics and Knowledge (LAK22)*.

Bannigan, K., & Moores, A. (2009). A model of professional thinking: Integrating reflective practice and evidence based practice. *Canadian Journal of Occupational Therapy, 76*(5), 342-350. https://doi.org/10.1177/000841740907600505

Baidoo-Anu, D., & Ansah, L. O. (2023). Education in the era of generative artificial intelligence (AI): Understanding the potential benefits of ChatGPT in promoting teaching and learning. *Journal of AI, 7*(1), 52-62. http://dx.doi.org/10.2139/ssrn.4337484

Brandt, R. M., & Perkins Jr, H. V. (1973). 12 Observation in Supervisory Practice and School Research. *ASCD-17948, 79*.

Brown, J., Hinze, S., & Pellegrino, J. W. (2008). Technology and formative assessment. *21st Century education, 2*, 245-255. http://dx.doi.org/10.4135/9781412964012.n77

Camacho, V. L., de la Guía, E., Olivares, T., Flores, M. J., & Orozco-Barbosa, L. (2020). Data capture and multimodal learning analytics focused on engagement with a new wearable IoT approach. *IEEE Transactions on Learning Technologies, 13*(4), 704-717. http://dx.doi.org/10.1109/TLT.2020.2999787

Cone, J. D. (1982). Validity of direct observation assessment procedures. *New Directions for Methodology of Social & Behavioral Science*.

Corbin, J. M., & Strauss, A. (1990). Grounded theory research: Procedures, canons, and evaluative criteria. *Qualitative sociology, 13*(1), 3-21. https://doi.org/10.1007/BF00988593

Dai, C. P., & Ke, F. (2022). Educational applications of artificial intelligence in simulation-based learning: A systematic mapping review. *Computers and Education: Artificial Intelligence, 100087*. http://dx.doi.org/10.1016/j.caeai.2022.100087

Deno, S. L. (1985). Curriculum-based measurement: The emerging alternative. *Exceptional children, 52*(3), 219-232. http://dx.doi.org/10.1177/001440298505200303

Dunphy, E. (2010). Assessing early learning through formative assessment: Key issues and considerations. *Irish Educational Studies, 29*(1), 41-56. https://doi.org/10.1080/03323310903522685

Erdemir, N., & Yeşilçınar, S. (2021). Reflective practices in micro teaching from the perspective of preservice teachers: teacher feedback, peer feedback and self-reflection. *Reflective Practice, 22*(6), 766-781. http://dx.doi.org/10.1080/14623943.2021.1968818

Grawemeyer, B., Mavrikis, M., Holmes, W., Gutiérrez-Santos, S., Wiedmann, M., & Rummel, N. (2017). Affective learning: Improving engagement and enhancing learning with affect-aware feedback. *User Modeling and User-Adapted Interaction, 27*, 119-158. http://dx.doi.org/10.1007/s11257-017-9188-z

Greenwood, C. R., Carta, J. J., Kamps, D., Terry, B., & Delquadri, J. (1994). Development and validation of standard classroom observation systems for school practitioners: Ecobehavioral Assessment Systems Software (EBASS). *Exceptional Children, 61*(2), 197. http://hdl.handle.net/1808/10954

Griffiths, M., & Tann, S. (1992). Using reflective practice to link personal and public theories. *Journal of Education for Teaching, 18*(1), 69-84. http://dx.doi.org/10.1080/0260747920180107

Halle, J. W., & Sindelar, P. T. (1982). Behavioral observation methodologies for early childhood education. *Topics in Early Childhood Special Education, 2*(1), 43-54. http://dx.doi.org/10.1177/027112148200200109

Han, C., & Lu, X. (2021). Interpreting quality assessment re-imagined: The synergy between human and machine scoring. *Interpreting and Society, 1*(1), 70-90. https://doi.org/10.1177/27523810211033670

Herman, J. L. (1992). *A practical guide to alternative assessment*. Association for Supervision and Curriculum Development, 1250 N. Pitt Street, Alexandria, VA 22314.

Holstein, K., McLaren, B. M., & Aleven, V. (2019). Co-designing a real-time classroom orchestration tool to support teachers–AI complementarity. *Journal of Learning Analytics, 6*(2), 27–52. http://dx.doi.org/10.18608/jla.2019.62.3

Holstein, K., Yu, Z., Sewall, J., Popescu, O., McLaren, B. M., & Aleven, V. (2018). Opening up an intelligent tutoring system development environment for extensible student modeling. *In Artificial Intelligence in Education: 19th International Conference*, *AIED 2018*, London, UK, June 27–30, 2018, Proceedings, Part I 19 (pp. 169-183). Springer International Publishing. https://doi.org/10.1007/978-3-319-93843-1_13

Ihantola, P., Ahoniemi, T., Karavirta, V., & Seppälä, O. (2010). Review of recent systems for automatic assessment of programming assignments. *Proceedings of the 10th Koli calling international conference on computing education research* (pp. 86-93). http://dx.doi.org/10.1145/1930464.1930480

Jia, M., Tang, L., Chen, B. C., Cardie, C., Belongie, S., Hariharan, B., & Lim, S. N. (2022, October). Visual prompt tuning. *In European Conference on Computer Vision* (pp. 709-727). Cham: Springer Nature Switzerland. https://doi.org/10.1007/978-3-031-19827-4_41

Jørnø, R. L., & Gynther, K. (2018). What constitutes an 'actionable insight' in learning analytics?. *Journal of Learning Analytics, 5*(3), 198-221. http://dx.doi.org/10.18608/jla.2018.53.13

Kabudi, T., Pappas, I., & Olsen, D. H. (2021). AI-enabled adaptive learning systems: A systematic mapping of the literature. *Computers and Education: Artificial Intelligence, 2,* 100017. https://doi.org/10.1016/j.caeai.2021.100017

Karavirta, V., Korhonen, A., & Malmi, L. (2006). On the use of resubmissions in automatic assessment systems. *Computer science education, 16*(3), 229-240. http://dx.doi.org/10.1080/08993400600912426

Kim, J., Lee, H., & Cho, Y. H. (2022). Learning design to support student-AI collaboration: Perspectives of leading teachers for AI in education. *Education and Information Technologies, 27*(5), 6069-6104. https://doi.org/10.1007/s10639-021-10831-6

Kirillov, A., Mintun, E., Ravi, N., Mao, H., Rolland, C., Gustafson, L., Xiao, T., Whitehead, S., Berg, A. C., Lo, W.-Y., Dollár, P., & Girshick, R. (2023). Segment anything. *arXiv preprint arXiv:2304.02643*. https://doi.org/10.48550/arXiv.2304.02643

Kubsch, M., Czinczel, B., Lossjew, J., Wyrwich, T., Bednorz, D., Bernholt, S., ... & Rummel, N. (2022). Toward learning progression analytics—Developing learning environments for the automated analysis of learning using evidence centered design. *Frontiers in Education, 7*. https://doi.org/10.3389/feduc.2022.981910

Kuniyoshi, K., & Kurahashi, S. (2020). Simulation of learning effects of adaptive learning. *Procedia Computer Science, 176*, 2164-2172. https://doi.org/10.1016/j.procs.2020.09.253

Lee, A. V. Y., Luco, A. C., & Tan, S. C. (2023). A Human-centric automated essay scoring and feedback system for the development of ethical reasoning. *Educational Technology & Society, 26*(1), 147-159. https://www.jstor.org/stable/48707973

Lee, G. G., & Zhai, X. (2023). NERIF: GPT-4V for Automatic Scoring of Drawn Models. *arXiv preprint arXiv:2311.12990*. https://doi.org/10.48550/arXiv.2311.12990

Lee, U., Jung, H., Jeon, Y., Sohn, Y., Hwang, W., Moon, J., & Kim, H. (2023a). Few-shot is enough: exploring ChatGPT prompt engineering method for automatic question generation in English education. *Education and Information Technologies*, 1-33. http://dx.doi.org/10.1007/s10639-023-12249-8

Lee, U., Han, A., Lee, J., Lee, E., Kim, J., Kim, H., & Lim, C. (2023b). Prompt Aloud!: Incorporating image-generative AI into STEAM class with learning analytics using prompt data. *Education and Information Technologies*, 1-31. http://dx.doi.org/10.1007/s10639-023-12150-4

Li, J., Li, D., Savarese, S., & Hoi, S. (2023). Blip-2: Bootstrapping language-image pre-training with frozen image encoders and large language models. *arXiv preprint arXiv:2301.12597*. https://doi.org/10.48550/arXiv.2301.12597

Li, X., Yan, L., Zhao, L., Martinez-Maldonado, R., & Gasevic, D. (2023, March). CVPE: A Computer Vision Approach for Scalable and Privacy-Preserving Socio-spatial, Multimodal Learning Analytics. *In LAK23: 13th International Learning Analytics and Knowledge Conference* (pp. 175-185). http://dx.doi.org/10.1145/3576050.3576145

Lim, E. M. (2023). The effects of pre-service early childhood teachers' digital literacy and self-efficacy on their perception of AI education for young children. *Education and Information Technologies*, 1-27. https://doi.org/10.1007/s10639-023-11724-6

Lin, B., Zhu, B., Ye, Y., Ning, M., Jin, P., & Yuan, L. (2023). Video-LLaVA: Learning United Visual Representation by Alignment Before Projection. *arXiv preprint arXiv:2311.10122*. https://doi.org/10.48550/arXiv.2311.10122

Liu, H., Li, C., Li, Y., & Lee, Y. J. (2023a). Improved baselines with visual instruction tuning. *arXiv preprint arXiv:2310.03744*. https://doi.org/10.48550/arXiv.2310.03744

Liu, H., Li, C., Wu, Q., & Lee, Y. J. (2023b). Visual instruction tuning. *arXiv preprint arXiv:2304.08485*. https://doi.org/10.48550/arXiv.2304.08485

Loughran, J. J. (2002). *Developing reflective practice: Learning about teaching and learning through modeling*. London: Routledge. https://doi.org/10.4324/9780203453995

Mao, J., Chen, B., & Liu, J. C. (2023). Generative artificial intelligence in education and its implications for assessment. *TechTrends*, 1-9. http://dx.doi.org/10.1007/s11528-023-00911-4

Marchisio, M., Barana, A., Fioravera, M., Rabellino, S., & Conte, A. (2018, July). A model of formative automatic assessment and interactive feedback for STEM. *In 2018 IEEE 42nd Annual Computer Software and Applications Conference* (COMPSAC) (Vol. 1, pp. 1016-1025). IEEE. http://dx.doi.org/10.1109/COMPSAC.2018.00178

Martínez-Maldonado, R., Yan, L., Deppeler, J., Phillips, M., & Gašević, D. (2022). Classroom analytics: Telling stories about learning spaces using sensor data. *In Hybrid learning spaces* (pp. 185-203). Cham: Springer International Publishing. https://doi.org/10.1007/978-3-030-88520-5_11

Moon, J., Lee, D., Choi, G. W., Seo, J., Do, J., & Lim, T. (2023). Learning analytics in seamless learning environments: a systematic review. *Interactive Learning Environments*, 1-18. http://dx.doi.org/10.1080/10494820.2023.2170422

Nazaretsky, T., Cukurova, M., & Alexandron, G. (2022, March). An instrument for measuring teachers' trust in AI-based educational technology. *In LAK22: 12th international learning analytics and knowledge conference* (pp. 56-66). https://doi.org/10.1145/3506860.3506866

Ndukwe, I. G., & Daniel, B. K. (2020). Teaching analytics, value and tools for teacher data literacy: A systematic and tripartite approach. *International Journal of Educational Technology in Higher Education, 17*(1), 1-31. http://dx.doi.org/10.1186/s41239-020-00201-6

Nian, Z. (2020) To Promote the Development of Teachers' Teaching Beliefs from Reflective Teaching. *Open Journal of Social Sciences, 8*, 120-126. https://doi.org/10.4236/jss.2020.811012

Nilsson, P., & Karlsson, G. (2019). Capturing student teachers' pedagogical content knowledge (PCK) using CoRes and digital technology. *International Journal of Science Education, 41*(4), 419-447. http://dx.doi.org/10.1080/09500693.2018.1551642

Olugbade, T. A., Bianchi-Berthouze, N., Marquardt, N., & Williams, A. C. D. C. (2018). Human observer and automatic assessment of movement related self-efficacy in chronic pain: from exercise to functional activity. *IEEE Transactions on Affective Computing, 11*(2), 214-229. http://dx.doi.org/10.1109/TAFFC.2018.2798576

OpenAI. (2022). Introducing Whisper. *OpenAI*. https://openai.com/research/whisper

OpenAI. (2023a). GPT-4 Technical Report. *arXiv preprint arXiv:2303.08774*. https://doi.org/10.48550/arXiv.2303.08774

OpenAI. (2023b). GPT-4V(ison) system card. *OpenAI*. https://openai.com/research/gpt-4v-system-card

OpenAI. (2023c). OpenAI DevDay. *OpenAI*. https://devday.openai.com/

OpenAI. (2023d). OpenAI API reference. *OpenAI*. https://platform.openai.com/docs/api-reference

Ouyang, F., & Jiao, P. (2021). Artificial intelligence in education: The three paradigms. *Computers and Education: Artificial Intelligence, 2*, 100020. https://doi.org/10.1016/j.caeai.2021.100020

Ouyang, F., Zheng, L., & Jiao, P. (2022). Artificial intelligence in online higher education: A systematic review of empirical research from 2011 to 2020. *Education and Information Technologies, 27*(6), 7893-7925. http://dx.doi.org/10.1007/s10639-022-10925-9

Pellegrino, J. W. (2006). Rethinking and redesigning curriculum, instruction and assessment: What contemporary research and theory suggests. *Commission on the Skills of the American Workforce, Chicago*, 1-15.

Pellegrino, J. (2023), "Introduction: Arguments in support of innovating assessments", in Foster, N. and M. Piacentini (eds.), *Innovating Assessments to Measure and Support Complex Skills*, OECD Publishing, Paris, https://doi.org/10.1787/534c6ae3-en

Pellegrino, A. M., & Gerber, B. L. (2012). Teacher reflection through video-recording analysis. *Georgia Educational Researcher, 9*(1), 1. https://doi.org/10.20429/ger.2012.090101

Peterson, G., & Elam, E. (2020). *Observation and assessment in early childhood education*. Zero Textbook Cost.

Phillips, H. N., & Condy, J. (2023). Pedagogical dilemma in teacher education: bridging the theory practice gap. *South African Journal of Higher Education, 37*(2), 201-217. https://hdl.handle.net/10520/ejc-high_v37_n2_a12

Prieto, L. P., Sharma, K., Dillenbourg, P., & Jesús, M. (2016, April). Teaching analytics: towards automatic extraction of orchestration graphs using wearable sensors. *In Proceedings of the sixth international conference on learning analytics & knowledge* (pp. 148-157).

Radford, A., Kim, J. W., Hallacy, C., Ramesh, A., Goh, G., Agarwal, S., Sastry, G., Askell, A., Mishkin, P., Clark, J., Krueger, G., & Sutskever, I. (2021, July). Learning transferable visual models from natural language supervision. *International conference on machine learning* (pp. 8748-8763). PMLR. https://proceedings.mlr.press/v139/radford21a.html

Resch, K., & Schrittesser, I. (2023). Using the Service-Learning approach to bridge the gap between theory and practice in teacher education. *International Journal of Inclusive Education, 27*(10), 1118-1132. https://doi.org/10.1080/13603116.2021.1882053

Resch, K., Schrittesser, I., & Knapp, M. (2022). Overcoming the theory-practice divide in teacher education with the 'Partner School Programme'. A conceptual mapping. *European Journal of Teacher Education*, 1-17. https://doi.org/10.1080/02619768.2022.2058928

Schön, D. A. (1992). *The reflective practitioner: How professionals think in action*. Routledge. https://doi.org/10.4324/9781315237473

Shermis, M. D., & DiVesta, F. J. (2011). *Classroom assessment in action*. Rowman & Littlefield Publishers.

Singh, T. (2023). Awesome-gpt4. *Github*. https://github.com/taranjeet/awesome-gpt4#gpt-4-vision

Slade, M. L., Burnham, T. J., Catalana, S. M., & Waters, T. (2019). The Impact of Reflective Practice on Teacher Candidates' Learning. *International Journal for the Scholarship of Teaching and Learning, 13*(2), 15. http://dx.doi.org/10.20429/ijsotl.2019.130215

Slater, S., & Baker, R. (2019). Forecasting future student mastery. *Distance Education, 40*(3), 380-394. http://dx.doi.org/10.1080/01587919.2019.1632169

Snodgrass Rangel, V., Bell, E. R., Monroy, C., & Whitaker, J. R. (2015). Toward a new approach to the evaluation of a digital curriculum using learning analytics. *Journal of Research on Technology in Education, 47*(2), 89-104. http://dx.doi.org/10.1080/15391523.2015.999639

Tripp, T. R., & Rich, P. J. (2012). The influence of video analysis on the process of teacher change. *Teaching and teacher education, 28*(5), 728-739. https://doi.org/10.1016/j.tate.2012.01.011

Varatharaj, R. (2018). Assessment in the 21st century classroom: The need for teacher autonomy. *International Journal of Research and Innovation in Social Science (IJRISS), 2*(6), 105-109.

Vera, D., & Castilleja Trejo, M. (2016). Using Video to Enhance Observational Assessment. *Dimensions of Early Childhood, 44*(2), 4-10. https://files.eric.ed.gov/fulltext/EJ1150271.pdf

Wang, J., Liu, Z., Zhao, L., Wu, Z., Ma, C., Yu, S., Dai, H., Yang, Q., Liu, Y., Zhang, S., Shi, E., Pan, Y., Zhang, T., Zhu, D., Li, X., Jiang, X., Ge, B., Yuan, Y., Shen, D., ... Zhang, S. (2023). Review of large vision models and visual prompt engineering. *Meta-Radiology, 100047*. https://doi.org/10.1016/j.metrad.2023.100047

Wang, Q., Rose, C. P., Ma, N., Jiang, S., Bao, H., & Li, Y. (2022). Design and application of automatic feedback scaffolding in forums to promote learning. *IEEE Transactions on Learning Technologies, 15*(2), 150-166. http://dx.doi.org/10.1109/TLT.2022.3156914

Worsley, M., Anderson, K., Melo, N., & Jang, J. (2021). Designing analytics for collaboration literacy and student empowerment. *Journal of Learning Analytics, 8*(1), 30-48. http://dx.doi.org/10.18608/jla.2021.7242

Xie, Y., Zhou, L., Dai, X., Yuan, L., Bach, N., Liu, C., & Zeng, M. (2022). Visual clues: Bridging vision and language foundations for image paragraph captioning. *Advances in Neural Information Processing Systems, 35*, 17287-17300.

Zhang, J., Yin, Z., Chen, P., & Nichele, S. (2020). Emotion recognition using multi-modal data and machine learning techniques: A tutorial and review. *Information Fusion, 59*, 103-126. https://doi.org/10.1016/j.inffus.2020.01.011

# Appendix

**Appendix 1 - Video-prompt pair**

*Pairs of Video and matching rubric prompt*

| Video Name | One Frame Image of Video (Blurred) | Prompt |
|---|---|---|
| Taekwondo forward roll | 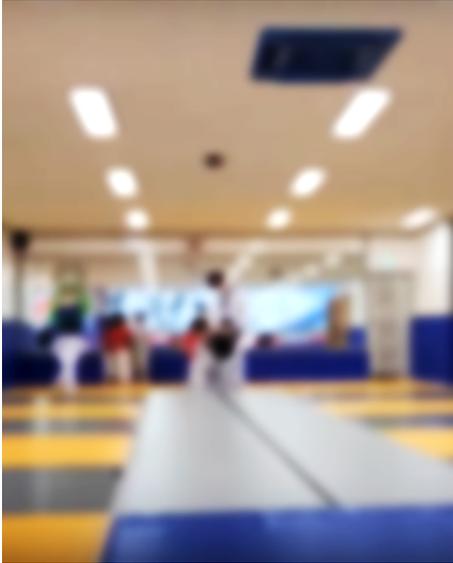 | 1. Place both hands on the ground: Start by placing both of your hands on the ground.<br>2. Curl your body into a round shape: Next, curl your body to form a round shape. This helps in creating momentum for the roll.<br>3. Initiate the roll: Gently push off with your legs. As you roll, make sure the back of your head, shoulders, back, hips, and finally your feet touch the ground in that order. This sequential contact is crucial for a smooth and safe roll.<br>4. Maintain correct body curvature: Lastly, be careful not to roll with your body too tightly curled or with your back too straight, as if lying down. Maintaining a proper curvature of the body is essential for a safe and effective forward roll.<br>5. Avoid improper body alignment: Be cautious not to lean your body to the right or left side while rolling. This can cause an uneven roll and increase the risk of injury.<br>6. Stand up: After completing the roll, use your feet to stand up. It's important to ensure that it's the back of your head that touches the ground, not the crown. Touching the ground with the crown of your head can lead to injury. |
| Class Demonstration | 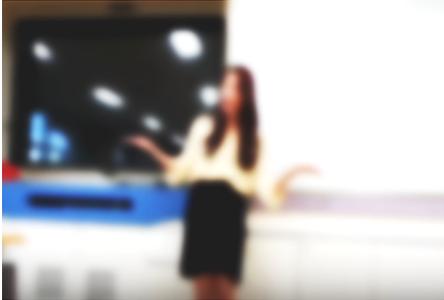 | ● Are the learning objectives set reasonably and stated clearly?<br>● Has motivational activity related to the learning objectives been effectively carried out?<br>● Are the elements of the learning content presented systematically and effectively?<br>● Were appropriate teaching methods used that align with the learning objectives and content?<br>● Were suitable questions and feedback provided that align with the learning objectives and content? |

|  |  | <ul><li>Were appropriate media and materials used for the learning activities, and was board writing done effectively?</li><li>Are learners actively engaged in the learning activities?</li><li>Was an appropriate evaluation conducted to confirm the achievement of the learning objectives?</li><li>In case of changes in the classroom situation, was the lesson plan adjusted smoothly if needed?</li><li>Were language and non-verbal actions used appropriately for communication with the learners?</li></ul> |
| --- | --- | --- |
| Fire extinguisher usage drill | 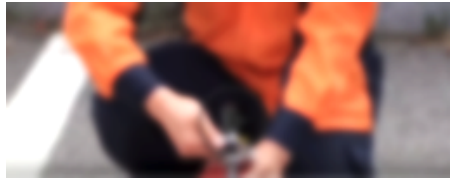 | 1. Approach the fire extinguisher and grasp the neck part, not the handle.<br>2. Remove the safety pin while holding the extinguisher on the ground.<br>3. Grab the lower part of the handle and proceed to the fire scene.<br>4. Stand 2-3 meters away from the fire.<br>5. Face away from the wind.<br>6. Hold the nozzle, press the handle, and sweep like using a broom. |
| Forklift drill | 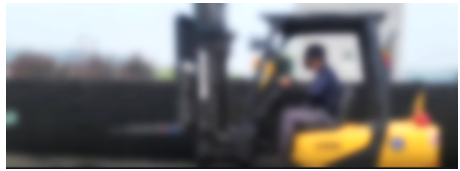 | <ul><li>Upon the start signal, lift the fork and cross the starting line.</li><li>Safely insert the fork into the pallet placed on the drum in the work area, then lift.</li><li>Ensure not to drop the drums or cargo.</li><li>Drive along the course following the forward indicators.</li></ul> |

**Appendix 2 - Usability Test Table**

*Findings from Qualitative Coding*

| **Themes** | **Codes** | **Implications** | **Related quotes** |
| --- | --- | --- | --- |
| 1. Usability of current VidAAS tool | 1.1. Advantages | 1.1.1. High rate of correct evaluation in the behavioral (psychomotor) domain | *"One of the things that I'm happy about is that there's a detailed explanation of exactly why it gave this score. (Translated)"* |
|  |  | 1.1.2. Comprehensive explanation for the evaluation |  |
| | 1.2. Disadvantages | 1.2.1. Limited video length and video processing latency | *"I've been putting videos in, one by one, and it takes a lot of time. (Translated)"* |
| | | 1.2.2. Unfeasible massive amount of assessment | |
| | | 1.2.3. Low rate of recognition precision | |
| 2. Potential benefits and improvement points of future VidAAS tool | 2.1. Benefits | 2.1.1. Potential of observing and evaluating in authentic contexts | *"I think we introduced process-based assessment because we don't want to do a one-time tense assessment, but we want to lower the psychological barrier to assessment and see them more naturally express their skills (and real-time video assessment can be helpful for that…) (Translated)"* |
| | | 2.1.2. Potential to mitigate assessment anxiety | |
| | | 2.1.3. Potential to enhance the range of evaluation methodologies | |
| | | 2.1.4. Potential to enhance evaluation accuracy and objectivity(fairness) | |
| | | 2.1.5. Potential to comprehend students' innate learning process with real-time analysis | |
| | | 2.1.6. Potential to simultaneous massive evaluation | |
| | | 2.1.7. Potential to provide feedback, scaffoldings and encouragement for students as a self-evaluation instrument | |
| | 2.2. Improvements | 2.2.1. Considering comprehensive evaluation of past cumulative observations | *"I think when human teacher evaluates someone, they can't ignore the progress and the effort and all of that, …. I think that's inevitably going to be reflected in the way that the child gets here. (Translated)"* |
| | | 2.2.2. The ability to consider context comprehensively to for the evaluation of cognitive and affective domain | |
| | | 2.2.3. Finding out extrinsic indicators for observations that can infer cognitive activities | |
| 3. Considerations of usage in relation to teacher analytics | 3.1. Reflection in | 3.1.1. A tool for teachers to do learning analytics in class | *"But in a case like this, if I can analyze all students instead of just one, then I'll keep the whole thing in focus and trace it, which would be helpful. (Translated)"* |
| | | 3.1.2. A tool to facilitate real-time, process-oriented assessment | |
| | 3.2. Reflection on | 3.2.1. Watching a recording and reflecting on the lesson | *"I think VidAAS could give students a score of the solving process.... When I give feedback to the parents, I think it would be good to show what is occurring in the problem solving process. I think that would be good to improve the teacher's profession. (Translated)"* |
| | | 3.2.2. Looking back at student's or teacher's performance | |
| | | 3.2.3. Data-driven decision-making | |
| | 3.3. Doubts (Limitations) | 3.3.1. Undetailed understanding of qualitative factors | *"It's just utilizing the tool, but if you give students VidAAS to use, it's going to help them build some skills. And the teacher is going to be helped rather than empowered. (Translated)"* |
| | | 3.3.2. Helping students improve their skills as a feedback tool, but not helping teachers improve their skills | |
| 4. Requirements for effective and ethical utilization | 4.1. Usage implications | 4.1.1. Informed of the need for contextualized interpretation of human teachers | *"I think it's just a tool to help teachers' observations. I mean, after staying together with a student for a month or so, I've gotten to know them… (Translated)"* |
| | | 4.1.2. Considering developmental level of the user | |
| | | 4.1.3. Contextualized usage process (protocol) in detail | |
| | | 4.1.4. Choosing the right device for the usage purposes | |
| | 4.2. Design principles | 4.2.1. Offer further assistance in establishing the rubric. | *"If the digital textbook platform in which teachers can take and aggregate video of their students is established, I think this kind of trinity makes it possible to use this tool more effectively. (Translated)"* |
| | | 4.2.2. Create accessibilities of different programmes and devices for various usage purposes (e.g. digital textbooks, smart glasses). | |
| | | 4.2.3. Create distinct interfaces to teachers and students | |
| | | 4.2.4. Archive video footage and assessment records to enable the examination of the longer-term performance improvement. | |
| | | 4.2.5. Reduce latency of video processing to enable real time analysis. | |
| | | 4.2.6. Enable massive evaluation for multiple students. | |
| | | 4.2.7. Analyze long and sophisticated video context to evaluate high-order thinking. | |
| | 4.3. Ethical guidelines | 4.3.1. Replacement of identifiable personal information | *"If you're going to utilize this app, I think it would be nice to have a notice at the beginning of each school year that says, 'We may use your child's video'. (Translated)"* |
| | | 4.3.2. Thorough functional readiness for data security | |
| | | 4.3.3. Consent to use of personal information | |
| 5. Diverse perspectives regarding AI | 5.1. Attitudes toward AI | 5.1.1. Satisfaction with the current level and a positive perception for future advances | *"I was a little curious about how much AI can evaluate the class situation, but I felt like it understood it at a fairly high level, so I was satisfied here for now... (Translated)"* |
| | | 5.1.2. Perspective of AI that it cannot be seen as a person | |
| | 5.2. AI-human collaboration | 5.2.1. Perspectives of AI (whether as a meer tool or a collaborative entity) | *"I think it's a tool, and the only time I would call it collaboration is when you have equal knowledge, equal capabilities and attitudes, and different opinions. (Translated)"* |
| | | 5.2.2. AI as a distributed cognition | |
| | 5.3. Perceptions of AI-assisted assessment | 5.3.1. Positive attitudes with a high level of understanding of the rubric | *"It's a little bit unclear how much it understands the learning objectives or social communication skills, so I'm* |
| | | 5.3.2. Negative attitudes with a distrust of AI | |

*a little bit skeptical that it can be trusted. (Translated)"*

**Appendix 3 - Design principles for VidAAS**

Design principles consist of three components; User Experience, Design, and Functions. First, for the User Experience; 1-1) Offer further assistance in establishing the rubric. To mitigate user confusion throughout the rubric setup process, it is imperative to break down the stages involved in setting up the rubric and provide exemplar rubrics throughout the various domains. 1-2) Create accessibilities to different platforms for various usage purposes. In order to accommodate the diverse usage needs of VidAAS, it is imperative to develop a scalable design for VidAAS that is compatible with other educational technology programs and devices. Second, for the Design perspective; 2-1) Create distinct interfaces tailored to the specific needs of teachers and students. Learners should get not only factual analysis but also encouraging words, while teachers should have access to interfaces that are optimized for managing numerous learners. 2-2) Enable the examination of the longer-term performance improvement. The users should be allowed access to stored video footage and automated evaluation records, and it would also be beneficial for users to receive an analysis of the progression processes. Lastly, for the Functions perspective; 3-1) Enable real-time analysis and evaluation. It is necessary to decrease the latency of video processing and enable the transmission and analysis of real-time video. 3-2) Enable concurrent mass evaluation for multiple students. 3-3) Evaluate high-order thinking and emotional capabilities. By making it possible to consider the long and sophisticated context, it would be feasible to enhance the assessment of cognitive and affective domains

*Design principles for VidAAS*

| Components | Principles and specific instructions |
|---|---|
| 1. User Experience | 1.1. Offer further assistance in establishing the rubric. |
| | 1.1.1. Develop the process of establishing the rubric into distinct steps. |
| | 1.1.2. Provide an example rubric for setting up a custom rubric. |
| | 1.2. Create accessibilities to different platforms for various usage purposes. |
| | 1.2.1. Make it work with various edutech devices (e.g., digital textbooks). |
| | 1.2.2. Make it work with wearable devices (e.g., smart glasses). |
| 2. Design | 2.1. Create distinct interfaces tailored to the specific needs of teachers and students. |
| | 2.1.1. Make sure the feedback provided to learners should be motivating. |
| | 2.1.2. Make sure the interface presented to learners should be optimized for recording and uploading videos. |
| | 2.1.3. Make sure the interface provided to teachers should allow them to manage multiple students simultaneously. |
| | 2.2. Enable the examination of the longer-term performance improvement. |
| | 2.2.1. Make it possible for users to view evaluation progress through the archived video footage. |
| | 2.2.2. Make it possible for users to view evaluation progress through the automated assessment records. |
| | 2.2.3. Provide additional analysis of the progress of improvements. |
| 3. Functions | 3.1. Enable real-time analysis and evaluation. |
| | 3.1.1. Reduce the latency of video processing. |
| | 3.1.2. Enable real-time video transmission and its subsequent analysis. |
| 3.2. Enable concurrent mass evaluation for multiple students | |
| | 3.2.1. Enable simultaneous analysis of a substantial quantity of videos. |
| 3.3. Evaluate high-order thinking and emotional capabilities. | |
| | 3.3.1. Make it possible to consider the long and sophisticated context. |